\documentclass{article}
\usepackage[left=4cm, right=4cm, top=4cm]{geometry}
\usepackage[utf8]{inputenc}
\usepackage{graphicx,subcaption}
\usepackage{amsmath}
\usepackage{booktabs}
\usepackage{multirow}
\usepackage{longtable}
\usepackage{hyperref}
\usepackage[nottoc]{tocbibind}
\usepackage{arydshln}
\usepackage[table]{xcolor}
\usepackage{pdfpages}

\makeatletter
\let\blx@rerun@biber\relax
\makeatother

\begin{document}

\begin{titlepage}

\newcommand{\HRule}{\rule{\linewidth}{0.5mm}} 

\center 


\HRule \\[0.5cm]
{ \huge \bfseries Structure and Automatic Segmentation of Dhrupad Vocal Bandish Audio}\\[0.5cm] 
\HRule \\[1cm]


\Large An R\&D project report submitted towards partial fulfillment of the requirements for the degree of \\[.25cm] \textbf{Master of Technology}\\[0.5cm] 

\Large{ By \\ \vspace{0.5cm} \textbf{M A Rohit\\ 
Roll Number: 183076001}\\
\vspace{1cm}
Under the supervision of:\\ \vspace{0.5cm}
\textbf{Prof. Preeti Rao}}\\ 
\vspace{1cm}

\center{\includegraphics[scale=0.15]{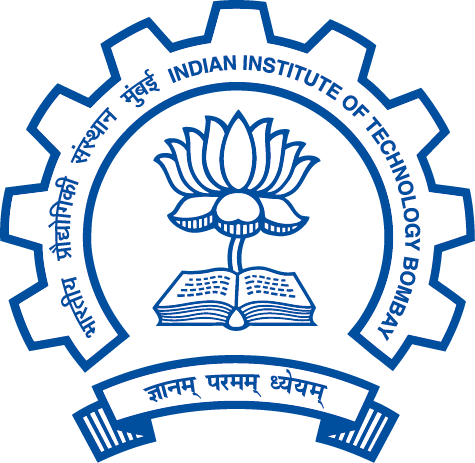}\\[0.5cm]}

\Large{Department of Electrical Engineering\\
Indian Institute of Technology Bombay\\
Mumbai 400076, India\\}
\vspace{1cm}

\Large{May 2020}

\vfill 

\end{titlepage}
\pagebreak

\begin{abstract}
\normalsize
A Dhrupad vocal concert comprises a composition section that is interspersed with improvised episodes of increased rhythmic activity involving the interaction between the vocals and the percussion.  Tracking the changing rhythmic density, in relation to the underlying metric tempo of the piece, thus facilitates the detection and labeling of the improvised sections in the concert structure.  This work concerns the automatic detection of the musically relevant rhythmic densities as they change in time across the \textit{bandish} (composition) performance. An annotated dataset of Dhrupad bandish concert sections is presented. We investigate a CNN-based system, trained to detect local tempo relationships, and follow it with temporal smoothing. We also employ audio source separation as a pre-processing step to the detection of the individual surface densities of the vocals and the percussion. This helps us obtain the complete musical description of the concert sections in terms of capturing the changing rhythmic interaction of the two performers.
\end{abstract}

\pagebreak

\tableofcontents

\pagebreak

\section{Introduction}
Dhrupad is one of the oldest forms of North Indian classical vocal music. A typical Dhrupad concert setting comprises a solo vocalist or vocalist duo as the lead and a pakhawaj player for the percussion accompaniment, with a tanpura in the background for the harmonic drone\cite{bonnie_wade}. A Dhrupad performance lasts for over an hour and consists of an elaborate, unaccompanied \textit{raga alap} followed by a composed piece, the \textit{bandish}, performed along with the percussion instrument. The bandish is not only presented as it is but also used as a means for further rhythmic improvisation (\textit{laykari}), where the vocalist sings the syllables of the bandish text at different rhythmic densities and in different patterns \cite[chapter~10]{clayton}. All the while, the pakhawaj accompaniment is either playing a basic pattern (\textit{theka}) of the metric cycle (\textit{tala}), a rhythmic improvisation to match the vocalist’s improvisation, or a free solo improvisation while the vocalist presents the lines of fixed composition or of the refrain \cite[p.~111]{clayton}. \\

\begin{figure}[!h]
\centering
\includegraphics[trim=50 250 90 100, clip, scale=0.7]{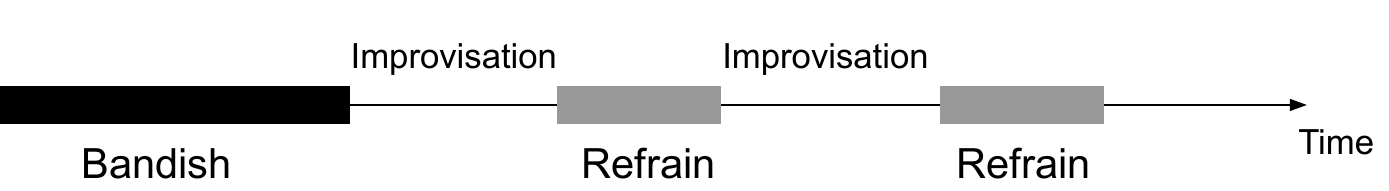}
 \caption{The structure of a bandish performance - vocalist's perspective \cite{clayton}}
\label{fig:concert_structure}
\end{figure}

Figure~\ref{fig:concert_structure} depicts the structure of a bandish performance from the perspective of the vocalist. The intermediate refrain portions are the un-improvised sections where the artist sings a line of the bandish or a portion of it before diving back into another spell of improvisation.\\

A key difference between Khyal and Dhrupad bandish performances is the relationship between the playing of the lead and accompanying percussion instrument. In the former, artists generally take turns improvising while the other provides the basic rhythm or the refrain, whereas in the latter, in addition to this, there are sections where both improvise together. A complete segmentation of a Dhrupad bandish performance would thus involve providing rhythmic descriptions of un-improvised and improvised sections pertaining to each - the vocals and the pakhawaj.\\ 

A few other levels of segmentation could also be realised on this form of music. One could be aimed at identifying sections of un-improvised and improvised music. And with the bandish text itself consisting of sections placed in different portions of an octave, each one with a couple of lines, another level of segmentation could be based on which of these is being presented or improvised upon by the vocalist. While the former segmentation could, to some extent, be defined by changes in rhythmic attributes, the latter would be based more on melodic characteristics. There could also be changes in timbral attributes correlated with the nature of articulation accompanying the various rhythmic or melodic changes, that could help with each kind of segmentation.\\

The goal of this work is to develop automatic methods for the structural segmentation of the Dhrupad bandish concert based only on the tempo and the relationships of the rhythmic densities of the individual instruments. Given that vocal onsets are acknowledgedly difficult to detect (even in isolated vocals due to the diversity inherent to singing), we turn to alternate methods for the direct estimation of the local rhythmic density. Advances in deep learning have led to the development of methods that treat the estimation of the predominant tempo from the raw audio spectral representation as a classification task \cite{schreiber-ismir18,schreiber-smc19,mtl-beat-tempo-19}. 
We explore a similar approach for our task of estimating the changing surface tempo or rhythmic density across a concert audio. In view of the significant improvements reported in audio source separation in recent years, we also consider the use of source separation followed by tempo estimation for the constituent instruments (vocals and pakhawaj) in order to give a more complete description of each section.\\


The chief new contributions of this work are as follows: (i) a dataset of tempo markings and rhythmic density based structural segmentation annotations for Dhrupad bandish concerts, (ii) adapting a state-of-the-art tempo estimation method to the task of estimating local surface tempo or rhythmic density of the polyphonic mix, and (iii) the use of source separation to extend this to each instrument(vocals and pakhawaj) to eventually obtain a musically relevant segmentation of bandish concerts with section labels defined in terms of the rhythmic density inter-relationships.

\section{Background}\label{sec:background}
Compositions in Hindustani music are sung at a tempo in one of roughly three broad ranges - \textit{vilambit} (10.4-60 BPM), \textit{madhya} (40-175 BPM) or \textit{drut} (170-500 BPM) \cite[p.~86]{clayton}. This tempo is determined by the interval between the \textit{matras} of the tala (a cyclic pattern of beats) that the composition is set to and is referred to as the metric tempo. The task of tracking the positions of the individual matras of the tala and thus identifying the cycle boundaries across a concert audio has been of research interest \cite{ajay14}. The metric tempo is fairly stable with only a gradual upward drift across the performance. However there are local variations in the rhythmic density of the singing or playing during what can be called episodes of improvisation, which constitute the surface rhythmic density or surface tempo. For the voice, this is calculated using the number of syllables or distinct notes uttered in a unit interval and for the pakhawaj, the number of strokes played in a unit interval \cite[p.~86]{clayton}. The surface tempo is found to generally be an integer multiple (ranging between 2 and 16) of the underlying metric tempo and we use the term `surface tempo multiple'(\textit{lay} ratio) to refer to this integer. Henceforth, we use the abbreviations m.t., s.t. and s.t.m. to refer to the metric tempo, surface tempo and surface tempo multiple.\\

The various kinds of sections in a Dhrupad bandish performance are summarised in Table~\ref{tab:band_sections}, in terms of the activity of each instrument. In the un-improvised vocal sections, the s.t.m. of the bandish or refrain singing is 1. In the \textit{bolbaat} section, the vocal improvisation is in the form of a rhythmic utterance of syllables, usually at a higher s.t.m., but sometimes at an s.t.m. equal to 1 in the early stages of the performance. The surface density is constituted mainly by distinct syllable onsets, although the vowels in these syllables can sometimes span a couple of pulses. In the \textit{boltaan} section, on the other hand, the vocal s.t.m. is always higher than 1, and is realised mainly through rapid note changes and not distinct syllable utterances. Due to these different forms of vocalisation, accurate vocal onset detection can be quite challenging \cite{vowel-onset-det}. The s.t.m is therefore calculated in a \textit{bolbaat} section using the rate of syllable onsets, and in a \textit{boltaan} section from the rate of note changes, with gaps and vowel elongations at a pulse level in each case treated as rhythmic events.\\

\begin{figure}[!h]
\centering
\begin{subfigure}{\linewidth}
\centering
\includegraphics[scale=0.3]{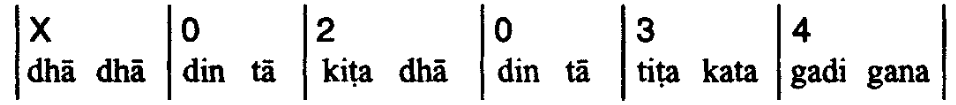}
\label{fig:chautal_basic}
\caption{}
\end{subfigure}

\begin{subfigure}{\linewidth}
\centering
\includegraphics[scale=0.3]{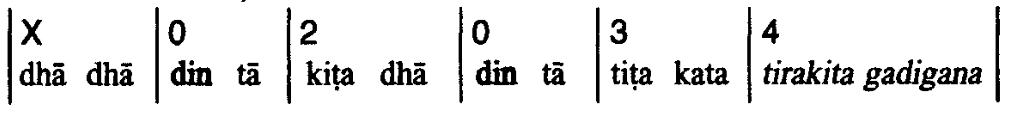}
\label{fig:chautal_div}
\caption{}
\end{subfigure}
\caption{The theka of \textit{Choutal}, a 12 matra cycle with 6 sub-divisions of 2 matras each. (a) Its basic form with 1 stroke per matra for the most part except on the last 4 matras, where it is 2 per matra. (b) A variation with 4 strokes on each of the last two matras \cite{clayton}.}
\label{fig:chautal}
\end{figure}

In the un-improvised vocal sections, the pakhawaj accompaniment is either in the form of an un-improvised theka or a free solo improvisation. During the theka-accompaniment, the s.t.m. of its realisation is not always the same, due to two reasons. One, the talas in Dhrupad have asymmetrical structures which already gives rise to portions of different densities within them (Figure \ref{fig:chautal} (a)). Two, the pakhawaj playing is sometimes in the form of a further divisive manipulation of the theka (Figure \ref{fig:chautal} (b)), akin to fillers in the tabla accompaniment in \textit{khyal} and instrumental \textit{gat} performances. During the pakhawaj solo section, the singing is un-improvised and the pakhawaj improvises freely at surface tempo multiples greater than 2. During improvised singing, the pakhawaj accompaniment is either in the form of the theka, or `synchronized accompaniment' (\textit{sath sangat}) in which the pakhawaj player tries to imitate the vocalist's rhythm.\\

Therefore, although the roles for each instrument are quite well defined, we see a lack of an exact correlation between the rhythmic density of an instrument's playing and the nature of playing (in terms of whether it is improvised or not).\\

\begin{table}
\centering
\begin{tabular}{@{}lll@{}}
\toprule
\textbf{Section} & \textbf{Vocal activity} & \textbf{Percussion activity}                                                  \\ \midrule
\multirow{2}{*}{Bandish / refrain} &
  \multirow{3}{*}[-0.7em]{\begin{tabular}[c]{@{}l@{}}Un-improvised singing \\ of bandish / refrain \\ at metric tempo\end{tabular}} &
  \multirow{2}{*}{\begin{tabular}[c]{@{}l@{}}Un-improvised playing \\ of theka\end{tabular}} \\
                 &                         &                                                                               \\ \cmidrule(r){1-1} \cmidrule(l){3-3} 
Pakhawaj solo    &                         & \begin{tabular}[c]{@{}l@{}}Solo / free rhythmic \\ improvisation\end{tabular} \\ \midrule
\multirow{2}{*}{Bolbaat / Boltaan} &
  \multirow{3}{*}[-1em]{\begin{tabular}[c]{@{}l@{}}Improvised singing \\ using the bandish text\end{tabular}} &
  \multirow{2}{*}{\begin{tabular}[c]{@{}l@{}}Un-improvised playing \\ of theka\end{tabular}} \\
                 &                         &                                                                               \\ \cmidrule(r){1-1} \cmidrule(l){3-3} 
Sath sangat &
   &
  \begin{tabular}[c]{@{}l@{}}Improvised playing \\ synchronised with vocals\end{tabular} \\ \bottomrule
\end{tabular}
\caption{The various sections of a Dhrupad bandish performance}
\label{tab:band_sections}
\end{table}

The m.t. and s.t. in this form of music have fairly objective definitions in terms of the performers' intentions and do not necessarily coincide with the `perceptual tempo'. And indeed as stated in \cite[p.~85]{clayton}, the perceived tempo at extreme values of the m.t. or s.t. may be quite different due to subdivisions at the lower end and grouping and accenting at the higher. For instance, the lowest m.t. measured by the author in \cite{clayton} was 10.4 BPM in a vocal concert, but the singing exhibited consistent subdivision of the matra at a rate equal to 4 times the m.t., even in the un-improvised portions, resulting in a perceptual tempo of nearly 42 BPM. Similarly, in the improvised sections, at high values of s.t., the tempo perceived can reduce to the \textit{vibhaag} rate (rate of occurrence of vibhaag boundaries) if one is familiar with the tala structure, or simply a rate equal to one of a half, third or fourth of the actual rhythmic density. To give a rough comparison with Western music, the two tempo levels m.t. and s.t. could be thought of as being similar to the tactus and tatum levels. In this regard, similar work on the estimation of different tempo levels can be found in \cite{klapuri05, peeters07, olivier19}, where tracking the various metrical levels is part of a system for tracking the predominant tempo as well as the meter of the audio. What is of more interest about these works, as also those of \cite{grosche09} and \cite{fhfwu11}, is the idea of tracking changing tempo across an audio and not only estimating a global tempo, which is a goal we share in this work.\\

Related work on structural segmentation for Hindustani classical music can be found in \cite{pr-tismir20, vinutha-ismir16, vinutha-ncc16, verma-icassp15}. The work in \cite{pr-tismir20} is regarding the segmentation of the initial unaccompanied alap portion of a Dhrupad vocal concert into the alap, jod and jhala sections. The methods are based on the changing nature of the energy, pulse clarity (salience), speed, and timbre of the vocals. In \cite{vinutha-ncc16, verma-icassp15}, the task of segmenting the unaccompanied,  and in \cite{vinutha-ismir16} the accompanied portion of instrumental concert audios consisting of a lead melodic instrument(sitar, sarod) and a tabla accompaniment, was addressed. Signal processing methods based on finding onsets followed by periodicity detection were made use of for tempo and rhythmic density estimation. Section boundaries were obtained with the help of a similarity detection matrix, using frame-level ACF vectors computed on the stream of detected onsets in \cite{vinutha-ismir16}, and using additional acoustic features and feature transformations in \cite{verma-icassp15}. Faced with the problem of two instruments playing together, differences in the instrument timbres were exploited to separate the plucked string and tabla onsets in \cite{vinutha-ismir16} to determine separately the metric and the surface tempo. Other source separation methods like HPSS\cite{aggelos12,elowson13} and PLCA\cite{chordia09} have also been used to obtain tempo estimates for individual sources, which are then combined together to refine the overall tempo estimate.\\

In this work we address the structural segmentation of the bandish section in Dhrupad vocal performances, which has not been attempted before. We propose to achieve this by first estimating the surface tempo using the CNN-based approach of \cite{schreiber-ismir18} with a modified architecture to predict it directly as a multiple of the metric tempo. To obtain the s.t.m. of each instrument, we make use of a pre-trained model provided by spleeter \cite{spleeter} that separates vocals from the accompaniment. We then detect section boundaries in a concert audio using changes in the estimated local s.t.m. values. 

\section{Dataset Description}\label{sec:}
To the best of our knowledge there is no existing dataset of tempo and segmentation related annotations for Dhrupad bandish performances. The dataset chosen for this work contains 14 concert audios - 8 from the Dunya corpus \cite{dunya} and the rest from publicly available recordings. 9 of the 14 are by the vocalist duo Gundecha brothers, and the others by Uday Bhawalkar. Each recording is of a single bandish performance by the vocals, accompanied by pakhawaj, with a tanpura in the background. The recordings are 8-15 minutes long and the total duration of the dataset is about 3 hours. The performances are not all in the same raga or tala with at least one composition in each of 4 distinct talas commonly found in Dhrupad. 7 more publicly available audios were partially annotated to balance the cross-validation dataset described in Section~\ref{sec:dataset_stats}. All the dataset details and annotations are made available, with the list of audios and links to their sources provided in the appendix. 

\subsection{Annotations}
Annotations are of (i) the \textit{sam} positions of the tala, i.e., the cycle boundaries, across the concert (ii) boundaries marking changes in the surface tempo multiple of each instrument and (iii) a label for each section in terms of the surface tempo multiple of each instrument. The annotations were marked by the author, who is a trained musician, using the relatively objective criteria described here.\\

Information about the tala and the number of matras was obtained from the metadata accompanying the recording. With this, the sam positions were then inferred either from the particular stroke of the pakhawaj or the syllable of the bandish refrain that appears on the sam in performance \cite{ross-ismir12}, or the number of matras elapsed since the previous sam. Although slight deviations are commonly observed in the metric tempo, large abrupt jumps do not occur. Hence, once a pass was made over the entire audio, the annotations were corrected at points of ambiguity to ensure coherence with adjacent sam markings.  The metric tempo was then calculated versus time, once for every cycle, by dividing the cycle duration by the number of matras in the tala. The tempo value at any point within a cycle is assumed to be the same.\\

A section boundary was marked whenever the rhythmic density of either instrument changed and the new density was maintained for at least a duration of 5s. As mentioned earlier, the surface tempo is typically related to the metric tempo as an integer multiple. Therefore every section was labelled with the surface tempo multiple of each instrument, determined by calculating the rate of events (syllables for the vocals and strokes for the pakhawaj) as a multiple of the metric tempo in the section. Pauses at the pulse level occurring between syllables or strokes were considered as musical events contributing to the surface tempo, while pauses longer than 5s were labeled as having no surface tempo. Sections with short and frequent changes in the density were labelled with the highest surface tempo exhibited.
The maximum of the vocal and pakhawaj surface tempo multiples was then added to the section label as the net surface tempo multiple denoting the overall level of rhythmic density. Denoting the overall activity as the maximum of the two is believed to be valid since the two instruments are usually in consonance with each other. 
The surface tempo in BPM versus time is obtained by multiplying the metric tempo and the surface tempo multiple. \\

\begin{figure}[!h]
\centering
\includegraphics[trim=15 100 15 20, clip, scale=0.7]{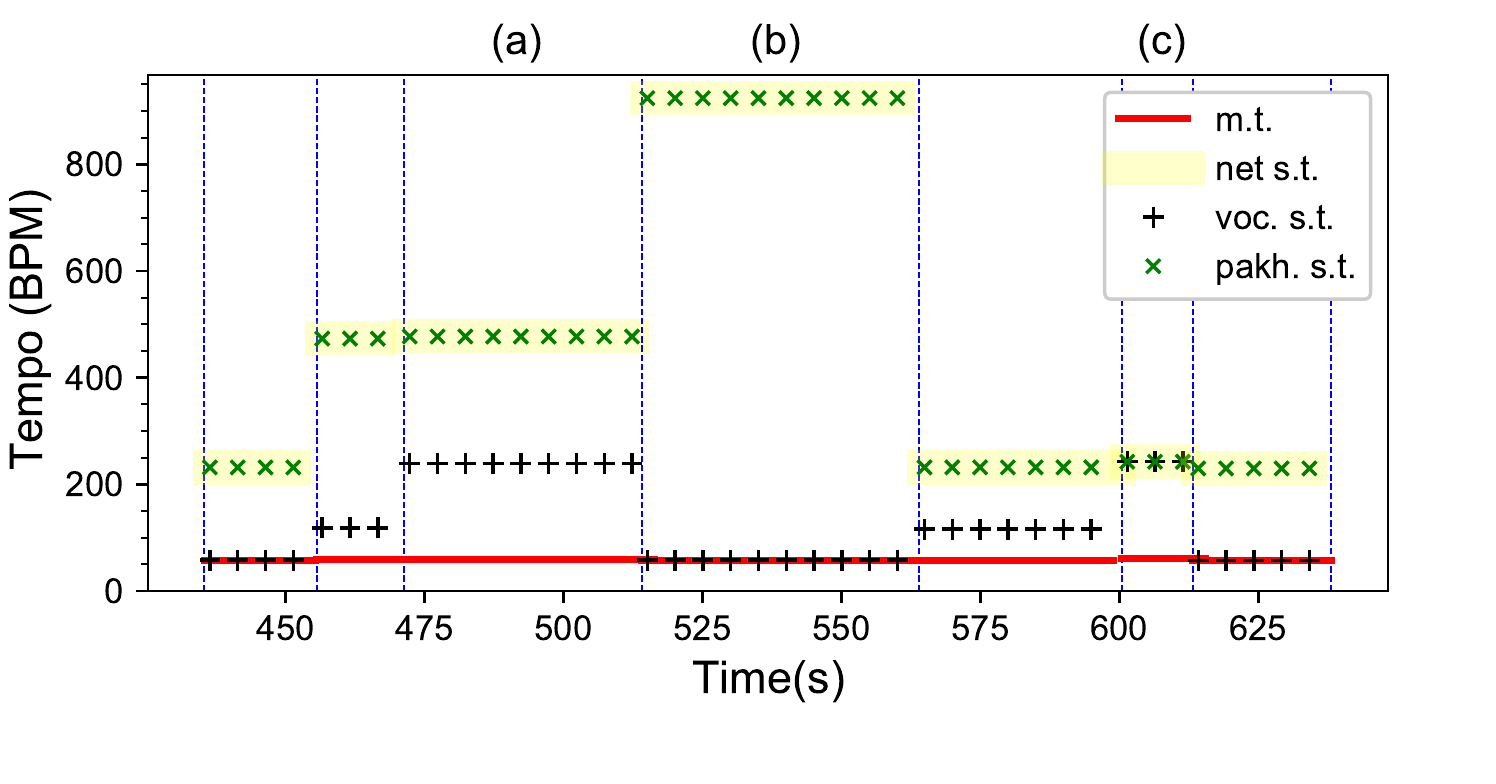}
 \caption{The ground truth metric tempo (m.t.) and surface tempo (s.t.) annotations for a portion of the audio \textit{GB\_AhirBhrv\_Choutal}. Vertical dashed lines indicate section boundaries.}
\label{fig:ann_vis}
\end{figure}

Figure~\ref{fig:ann_vis} is a visualisation of the annotations for a portion of a bandish performance by the Gundecha brothers (titled \textit{GB\_AhirBhrv\_Choutal} in the dataset). This roughly 4 minute long snippet captures a few different kinds of sections - (a) vocal s.t. at 4 times the m.t.($\sim$60 BPM) and pakhawaj at 8, (b) vocals at the m.t. and pakhawaj at 16 times - in each of these the net is due to the pakhawaj, and (c) both at 4 times, where the net is due to both.

\subsection{Dataset Statistics and Train-test Split}\label{sec:dataset_stats}
\begin{figure}[!h]
\centering
\begin{subfigure}{0.47\linewidth}
\centering
\includegraphics[trim=10 100 40 20, clip, scale=0.45]{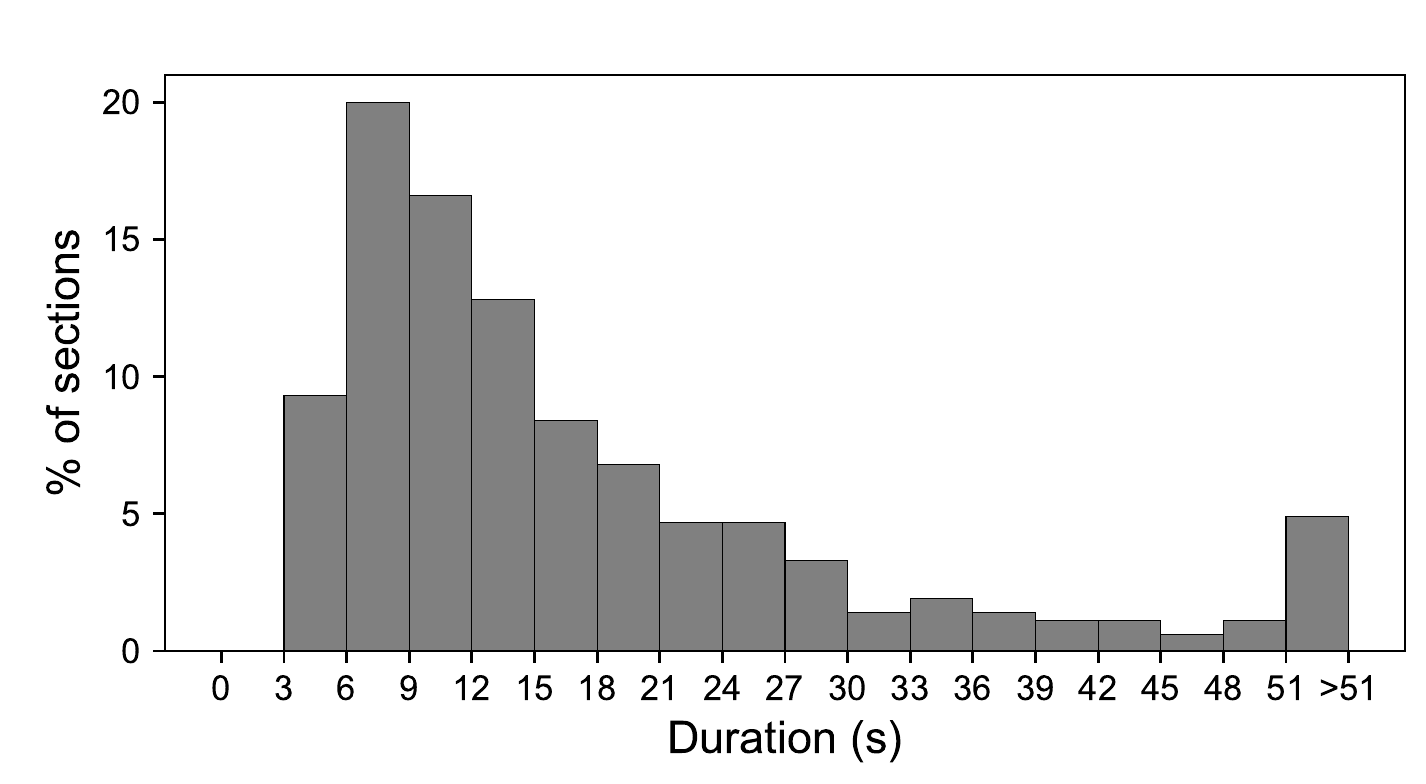}
\label{fig:data_stats_dur}
\caption{}
\end{subfigure}
~~~
\begin{subfigure}{0.47\linewidth}
\centering
\includegraphics[trim=20 100 30 20, clip, scale=0.45]{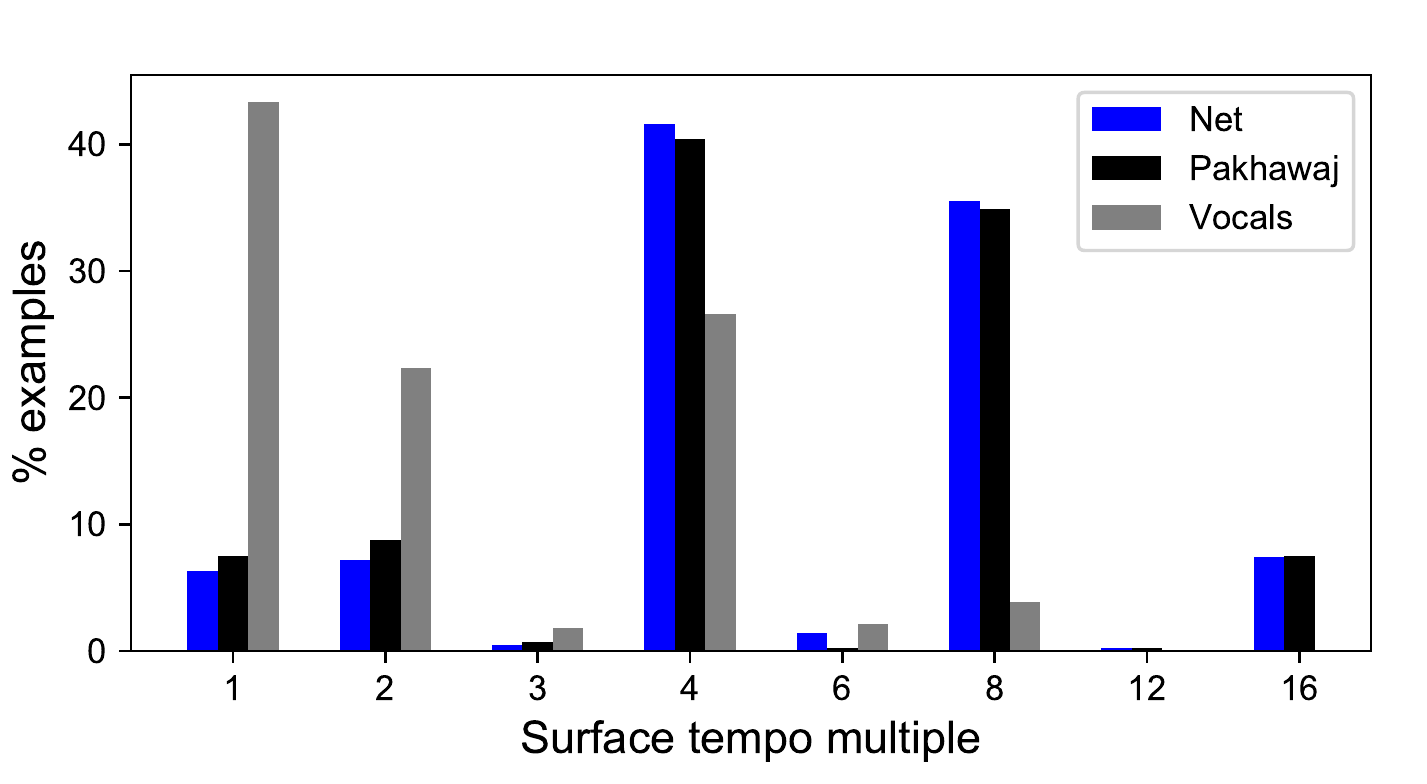}
\label{fig:data_stats_stm}
\caption{}
\end{subfigure}
\caption{Distributions of (a) section duration and (b) net, pakhawaj and vocal s.t.m. across our dataset.}
\label{fig:data_stats}
\end{figure}

Every annotated section is homogenous in the sense that the s.t. of each instrument remains the same throughout its duration. We therefore pool the sections from all the concert audios into a dataset for training and testing our methods, treating each section as an independent entity. The total number of sections comes up to 634 (593 from the completely annotated and the rest from the partially annotated audios), but they are not all of similar durations. Figure~\ref{fig:data_stats}~(a) shows the distribution of section durations over 3s intervals with a single bar at the end for values more than 51s. We see a peak in the 6-9s interval. With the goal of tracking the s.t. as it is changing across a performance, we need to perform tempo estimation on shorter examples from each section. The duration of these examples is set to be 8s since a higher value would give us no examples from the large number of sections that are only 6-9s long. Further, for the slowest tempo in the dataset of about 30 BPM, an 8s duration would contain at most 4 beats, fewer than which may not be sufficient for accurate tempo estimation.\\

The distribution of s.t.m. in the dataset for each instrument and the net is shown in Figure~\ref{fig:data_stats}~(b) in terms of the number of non-overlapping 8s examples (extracted from sections) available at each integer multiple. The dataset has a narrow m.t. range of 35 - 75 BPM, but the observed range of s.t. extends upto a large 990 BPM, due to the nature of the \textit{lay} ratios. For the pakhawaj, we find that the multiples 4 and 8 are more abundant than 1, 2 and 16, while the multiples 3, 6 and 12 are practically absent. For the vocals, 1, 2 and 4 are most represented and even though the multiples 3, 6 and 8 have a similar share, the sections for 8 were found to come from several concerts, while 3 and 6 were only found in a couple. We thus retain only the sections with s.t.m. values from the set \{1, 2, 4, 8, 16\}.\\

To manage the data imbalance, while generating the 8s training examples, the hop between consecutive examples is kept higher for sections belonging to the less populous s.t.m values. We also augment the dataset by time-scaling the audio of each section \cite{rubberband} using one or more factors in the range \{0.8, 0.84, 0.88, … 1.2\} (the s.t.m. label remains the same). Again, we generate more time-scaled versions for the less populous classes. The whole pool of examples is divided into three folds such that all the examples from a single section are assigned to the same fold, while ensuring that each fold has a similar distribution of examples across all the s.t.m. values.

\section{Methods}
We consider the recent CNN-based single-step tempo-estimation method from \cite{schreiber-ismir18} (denoted as tempo-cnn) for our work. After first examining the viability of the available pre-trained models, we attempt to train new models with the same architecture on our dataset, and then propose some modifications based on our observations. 

\subsection{Metric Tempo Estimation}
The m.t. of a Dhrupad bandish performance does not exhibit abrupt changes but does drift across a performance. Hence, we are interested in estimating it locally and tracking it versus time. With the m.t. range of our dataset being a subset of the pre-trained tempo-cnn output range, the model can be used as it is to observe the nature of its predictions. Upon obtaining estimates frame-wise at 0.5s hops and picking the output class with the highest confidence in each frame, it is found that the model almost always makes octave errors, which is to be expected since the m.t. in our case is not always the perceptual tempo that the model was trained to estimate. We fix these errors by constraining the predicted tempo to lie in the range of m.t. values in the dataset.\\

We do not attempt to train a new model for m.t. estimation and instead compare this method with a traditional, non-learning based approach using the algorithm from \cite{vinutha-ismir16}. A spectral flux based onset detector is used to obtain the onsets and the autocorrelation function is calculated on 12s long windows(similar to the input length for tempo-cnn) at 0.5s hops for values of lag upto 2s. The tempo candidates are constrained to be in the required range and an additional Viterbi smoothing step is used to penalise jumps and obtain a consistent estimate across a concert. We refer to this as the odf-acf method. We also note that the metrical cycle tracking work of \cite{ajay14} offers an alternative that can be investigated for m.t. estimation in future work.


\subsection{Surface Tempo Estimation}\label{sec:methods_stm}
The s.t. values in our dataset fall outside the tempo-cnn output range. And since the task is the very identification of the correct octave of the tempo, octave errors cannot be tolerated here, making the pre-trained tempo-cnn not directly applicable. If we are to try and re-train tempo-cnn on our dataset by increasing the output range, the huge size of the range presents a significant problem due to the resulting target class imbalance. Therefore, given that the s.t.m. is one of a set of integer values, we modify the task to predicting this multiple instead of the exact s.t. value. With the m.t. estimated separately, the s.t. can then be obtained by simply multiplying the m.t. with the estimated s.t.m. value.\\

\begin{table}[b]
\centering
\begin{tabular}{@{}llll@{}}
\toprule
& Layer                   & Dimensions        \\ \midrule
& Input                   & 40 x 400          \\
& (BN, Conv, ELU, DO) x3 & 16 x 1 x 5        \\
\arrayrulecolor{blue}\hline
\color{blue}{\vline}& AvgPool                 & 5 x 1           &\color{blue}{\vline}  \\
\color{blue}{\vline}& BN, MF Conv, DO & \begin{tabular}[c]{@{}l@{}}12x \{1x16, 1x32, 1x64, 1x96\}\end{tabular}& \color{blue}{\vline}\\
\color{blue}{\vline}& Concat, Conv            & 16 x 1 x 1  & \color{blue}{\vline}      \\
\arrayrulecolor{blue}\hline
& AvgPool                 & 1 x 400           \\
& BN, DO, FC, Softmax     & \# output classes \\ \arrayrulecolor{black}\bottomrule
\end{tabular}
\caption{Proposed model architecture, adapted from \cite{schreiber-ismir18} \& \cite{schreiber-smc19} (layers enclosed in blue constitute the MF module).}
\label{tab:model_arch}
\end{table}

\begin{figure}[!b]
\centering
\begin{subfigure}{0.6\linewidth}
\centering
\includegraphics[scale=0.45]{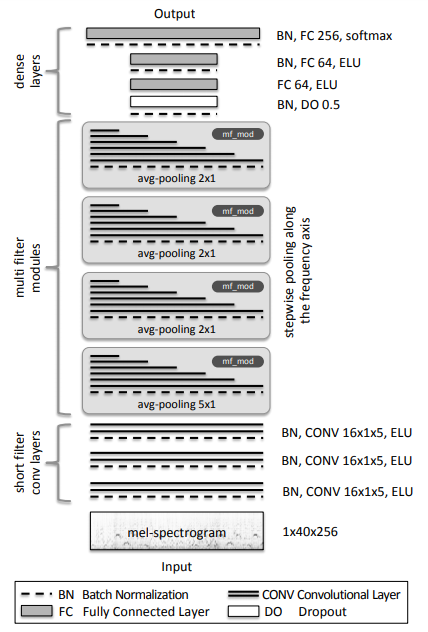}
\label{fig:tempo-cnn_arch}
\caption{}
\end{subfigure}
~
\begin{subfigure}{0.37\linewidth}
\centering
\includegraphics[scale=0.35]{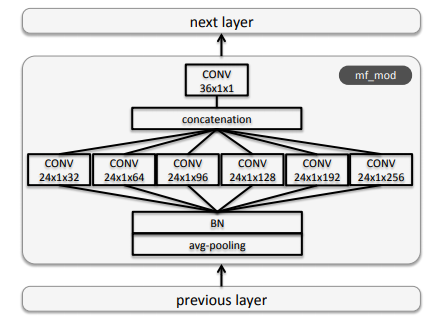}
\label{fig:mf_mod}
\caption{}
\end{subfigure}
\caption{(a) Architecture of the tempo-cnn network \cite{schreiber-ismir18} (b) The set of layers within a multi-filter (MF) module}
\label{fig:tempo-cnn_arch_mfmod}
\end{figure}

An attempt to train new models using the tempo-cnn architecture on our dataset by only reducing the dimensions of the final softmax layer does not turn out to be fruitful as the model quickly overfits due to its high capacity and the small size of our dataset. Notably, the main issues seem to be the high number of dense layers at the end and the large dimensions of the temporal filters in the multi-filter (MF) modules. After a series of simplifications with some inspiration from \cite{schreiber-smc19}, the architecture summarised in Table~\ref{tab:model_arch} is found to be promising. The original tempo-cnn architecture from \cite{schreiber-ismir18} is also shown for comparison (Figure \ref{fig:tempo-cnn_arch_mfmod} (a)). The 3 short filter conv layers in tempo-cnn are retained as they are and are followed by a single MF module (shown in Figure~\ref{fig:tempo-cnn_arch_mfmod} (b), and also marked by the blue rectangle in Table~\ref{tab:model_arch}), and a single dense (FC) layer. However, an addtional dropout layer is added after each short filter layer and after the MF Conv layer within the MF module. An AvgPool layer with the kernel oriented along time is also added before the dense layer.\\

The reduction in the number of dense layers and the addition of several dropout layers was found to be crucial in overcoming overfitting. To prevent too much information from getting cut-off due to the dropout, the p value is set to 0.1 in the first three conv layers, and 0.5 in the later ones. As for the MF Conv layer, fewer parallel filters and filter dimensions smaller than in tempo-cnn were found to make the network easier to train. However, to ensure adequate capacity, the number of filters in each layer was kept moderately high. The AvgPool layer at the end also helped reduce the number of trainable parameters in the dense layer and hence prevent overfitting.\\

This architecture was arrived at through a set of informal experiments which involved comparing the loss and accuracy curves during training and validation for a few different variations listed in Table~\ref{tab:model_arch_alt}. These variations included modifying the model depth and its capacity by changing the number of MF modules and the filter dimensions of the MF conv layer, examining the benefit of dropout in the early layers, and the benefit of the AvgPool layer before the dense layer. In each case, the rest of the architecture remained the same as in Table~\ref{tab:model_arch}, except that when the number of MF modules was more than 1, the AvgPool kernel size was 5x1 in the first and 2x1 in the subsequent MF modules. We note that model 2.a in Table~\ref{tab:model_arch_alt} corresponds to the proposed architecture. The resulting curves for all the cases are compared and discussed in Section~\ref{sec:results_stm}.

\begin{table}
\centering
\begin{tabular}{@{}llll@{}}
\toprule
Model     & \multicolumn{2}{c}{Description}                                   & \# parameters \\ \midrule
\multirow{2}{*}{1} & \multirow{2}{*}{\begin{tabular}[c]{@{}l|@{}}Filter dim. in MF Conv:\\ \{1x4, 1x6, 1x8, 1x12\}\end{tabular}}     & a. 1 MF module & 10,055 \\
          &                          & b. 3 MF modules                        & 22,823        \\ \\
\multirow{2}{*}{2} & \multirow{2}{*}{\begin{tabular}[c]{@{}l|@{}}Filter dim. in MF Conv:\\ \{1x16, 1x32, 1x64, 1x96\}\end{tabular}} & a. 1 MF module & 44,231 \\
          &                          & b. 3 MF modules                        & 125,351       \\ \\ 
3         & \multicolumn{2}{l}{Model 2.a without dropout in the early layers} & 44,231        \\ \\
4         & \multicolumn{2}{l}{Model 2.a without AvgPool before dense layer}  & 299,591       \\ \\
tempo-cnn & \multicolumn{2}{l}{As in \cite{schreiber-ismir18}, but with 5 output classes}       & 3,236,503     \\ \bottomrule
\end{tabular}
\caption{Alternatives to the proposed model architecture.}
\label{tab:model_arch_alt}
\end{table}



\subsection{Input Representation and Network Training}
Every 8s training example is transformed to a logarithmically scaled mel-filtered magnitude spectrogram, using the following parameters - a window size of 40ms, a hopsize of 20ms, and 40 mel filters over the band 20-8000 Hz, at a sampling rate of 16000 Hz. The input to the network is therefore a spectrogram of size 40 x 400 with the values normalized to lie in the range 0 - 1, and the target is one of 5 classes corresponding to the 5 s.t.m. values - \{1, 2, 4, 8, 16\}. The network is trained using categorical cross entropy loss on the examples from two folds, with the other fold as the validation set, for a maximum of 500 epochs. Training is carried out using the Adam optimizer with a learning rate of 1e-4 and a batch size of 32, and is halted early if the validation loss does not decrease for 50 epochs.

\subsection{Extension to Separated Sources}
The method described above operates on the original audios containing both the vocals and the pakhawaj, and therefore the estimated s.t.m. can only represent the overall level of rhythmic activity, without regards to the instrument responsible for it. 
Given our interest in estimating the s.t. of each instrument so that a more complete rhythmic description and thus the section boundaries in a concert can be obtained, the pre-trained 2-stems model released by spleeter \cite{spleeter} is used to separate the mixture audios into vocals and accompaniment and new models are trained to predict the s.t.m. for each. The same architecture as proposed above is used. The dataset of sections remains the same but the input examples are of the separated sources and the training and validation folds are generated again for each source, to balance the number of examples across the corresponding classes. The target classes for the pakhawaj are the same as earlier but those for vocals do not include the multiple 16. All the trained models and codes are made available.\\

\begin{figure}[!h]
\centering
\includegraphics[trim=75 0 0 0, clip, scale=0.5]{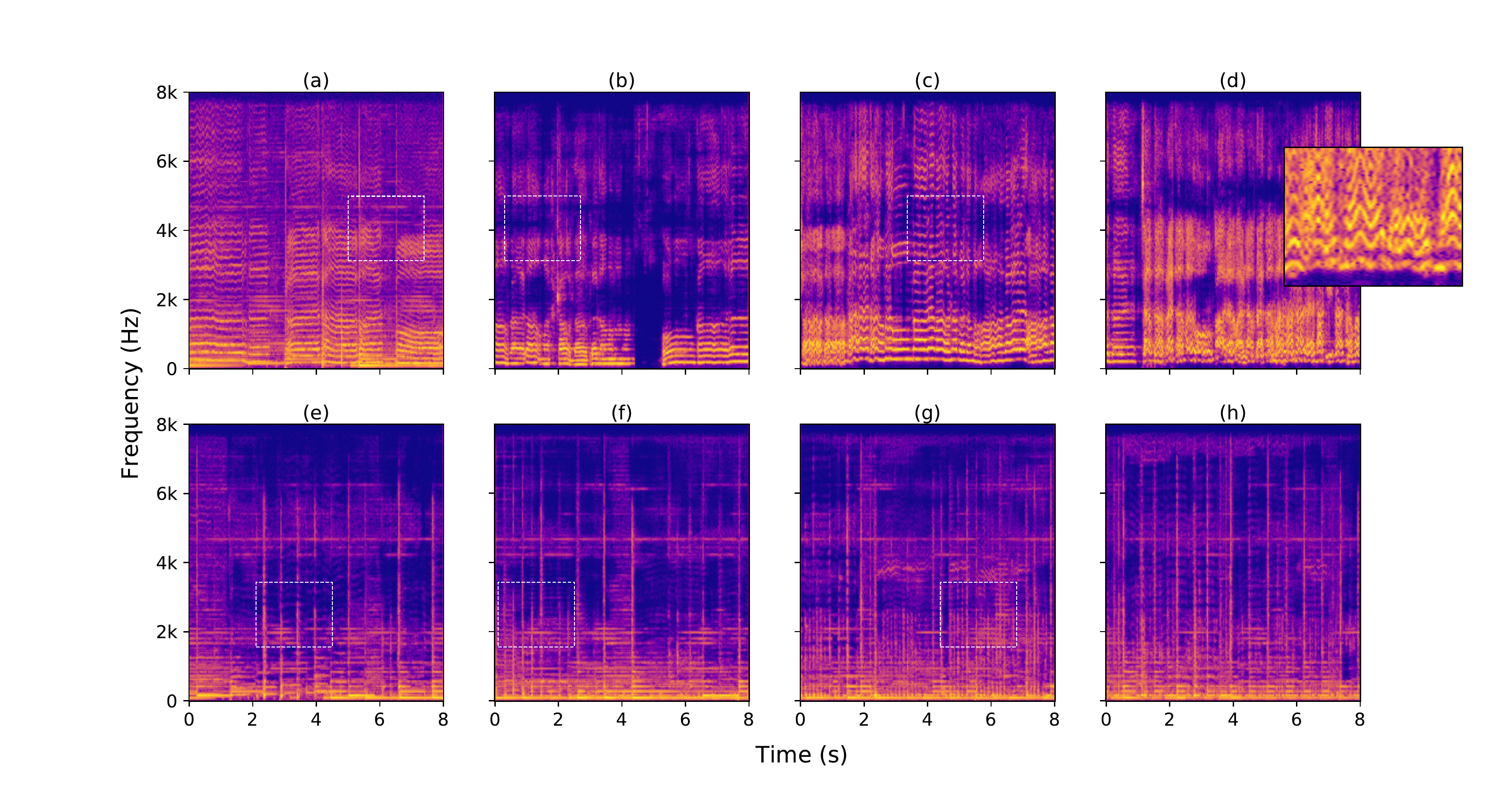}
 \caption{8s example spectrograms for each s.t.m. class. (a) Mixture audio (with s.t.m. equal to 1), (b) - (d) Separated vocals (s.t.m. values 2, 4, 8) and (e) - (h) Separated pakhawaj (s.t.m. values 2, 4, 8, 16). The dotted box (in white) of width ~2.5s is shown to highlight the differences in onset densities. The inset in (d) of width 1s highlights the distinct nature of vocalisation at s.t.m 8.}
\label{fig:specgram_examples}
\end{figure}

Figure~\ref{fig:specgram_examples} shows the spectrograms of some training examples (one for each s.t.m. class, taken from the concert \textit{GB\_AhirBhrv\_Choutal}. The plot (a) is of the mixture audio with s.t.m. 1, plots (b) - (d) are of the separated vocals, with vocal s.t.m. 2, 4 and 8 respectively, and plots (e) - (h) are of the separated pakhawaj, with pakhawaj s.t.m. 2, 4, 8 and 16. Also shown within the plots is a dotted rectangle of width ~2.5s, to help the reader estimate the number of strokes / syllables uttered in that time (the rectangle is positioned in a place where the density is most apparent). The m.t. in all the examples is between 50 and 60 BPM, so in each case, the number of strokes / syllables we expect to see within the rectangle is roughly equal to twice the s.t.m. value. This is fairly easy to see in plots (a) \& (e) - (g), where the tall, narrow,  vertical bursts of energy corresponding to the pakhawaj onsets help estimate the density. In plot (h) corresponding to s.t.m. 16, due to the extremely fast playing and an accenting at lower s.t.m. levels, the apparent density seems to match different s.t.m. values in different parts of the plot. However, there seems to be more persistent harmonic energy in the lower end of the spectrum, when compared to the other cases.\\

In the case of vocals, the s.t.m. is not as apparent. In (b), where the s.t.m. is 2, the narrow vertical regions with low energy marking syllable boundaries can be counted to estimate the s.t.m. value (these low energy regions are sometimes replaced by bursts of high energy in the higher portions of the spectrum when the syllable begins with a fricative sound). However in (c) where the s.t.m. is 4, boundaries are not always syllabic and are sometimes marked by abrupt changes in the notes. In (d), a 1s portion highlighting the distinct nature of vocalisation is inset. The density here is almost entirely from rapid pitch variations and not the utterance of distinct syllables. We expect to see 8 peaks in the melodic contour within the inset window. We see that we can indeed point to all the peaks except at the locations where the 3rd and 7th peaks must have been (there are pauses there instead).

\subsection{Boundary Detection and Section Labelling}
We aim to automatically identify sections in a concert by looking for abrupt changes in the s.t.m. values of each instrument across the concert duration. For this task only the completely annotated 14 concert audios are used. Estimates of the s.t.m. are obtained once every 0.5s using 8s long excerpts over the entire duration of each audio. While doing so, each excerpt is presented to that saved model out of the three from the 3-fold cross-validation procedure, to which no portion of the section that this excerpt lies in was presented as a training example, thus preventing any train-test leak. The output class with the highest confidence is taken as the s.t.m. estimate. This procedure is applied to the mixture and the source separated audios. A boundary is marked wherever the s.t.m. of either instrument changes, and the section label is the tuple of the three s.t.m. estimates. \\

We experiment with two methods for obtaining the three s.t.m. estimates. One, the three values are estimated independently, and we refer to this method as \texttt{seg1}. Here the net s.t.m. may not be equal to the higher of the other two (which should be true by definition). We thus report results using the model output for the net s.t.m. as well as by simply taking the maximum of the other two as the net s.t.m. value.
Two, to investigate whether using the relationship between the three s.t.m. values helps improve performance, instead of obtaining them independently, we pick that tuple of the three estimates in every frame which has the highest average classifier confidence value and in which the net s.t.m. is the maximum of the other two. We refer to this method as \texttt{seg2}. 
To reduce the number of false alarms, a post-processing step is used with each method to smooth the outputs by constraining the duration of a detected section to be at least 5s. This is implemented by removing the boundaries of any section that is shorter and replacing the label by that of the previous section. An alternative could be to replace the label with that of the next section, and this decision could be informed by observing the classifier output with the second highest confidence value.
We use the terms \texttt{seg1+}, \texttt{seg2+} to refer to the methods with the use of smoothing.\\

\section{Experiments and Results}\label{sec:results}

\subsection{Metric Tempo Estimation}
To evaluate m.t. estimation we calculate \textit{accuracy1} and \textit{accuracy2} (allowing for octave errors) values with a tolerance of 4\%  across each audio at a 0.5s frame-level and then average it across the dataset. The resulting scores using tempo-cnn without and with the tempo range constraint and the odf-acf method are shown in Table~\ref{tab:mt_acc}. We find that both the methods fare equally well and the simple fix of including a range constraint significantly improves \textit{accuracy1} for tempo-cnn  (except in cases where the prediction is an octave off but already in the correct range).\\

\begin{table}
\centering
\begin{tabular}{@{}lll@{}}
\toprule
Method                                                                    & Accuracy 1 & Accuracy 2 \\ \midrule
tempo-cnn                                                                 & 5.24       & 73.76      \\
\begin{tabular}[c]{@{}l@{}}tempo-cnn with\\ range constraint\end{tabular} & 71.61      & 74.75      \\ 

odf-acf                                                                 & 72.03        & 72.03      \\
\bottomrule
\end{tabular}
\caption{Metric tempo estimation accuracies (\%) at 4\% tolerance using tempo-cnn \cite{schreiber-ismir18} and the odf-acf method \cite{vinutha-ismir16}.}
\label{tab:mt_acc}
\end{table}

A closer look at concert-wise scores revealed that the accuracies were below 70\% in the same 4 (out of 14) concerts in both the methods, where most of the errors were due to the predicted value being either 1.5 or 0.75 times the actual m.t. value. While the tempo-cnn makes errors only in small portions of such concerts, in the other method, due to the imposed penalty on jumps, the predicted tempo was found to be incorrect over longer durations. Even so, what we take away from the overall results is that for most of the concerts, m.t. is estimated well across the entire duration despite the presence of sections where both the instruments are improvising and playing at different multiples of the m.t. (except perhaps when accents in the playing happen to favour a fractional level).

\subsection{Surface Tempo Estimation}\label{sec:results_stm}
\begin{figure}[!h]
\centering
\includegraphics[scale=0.4]{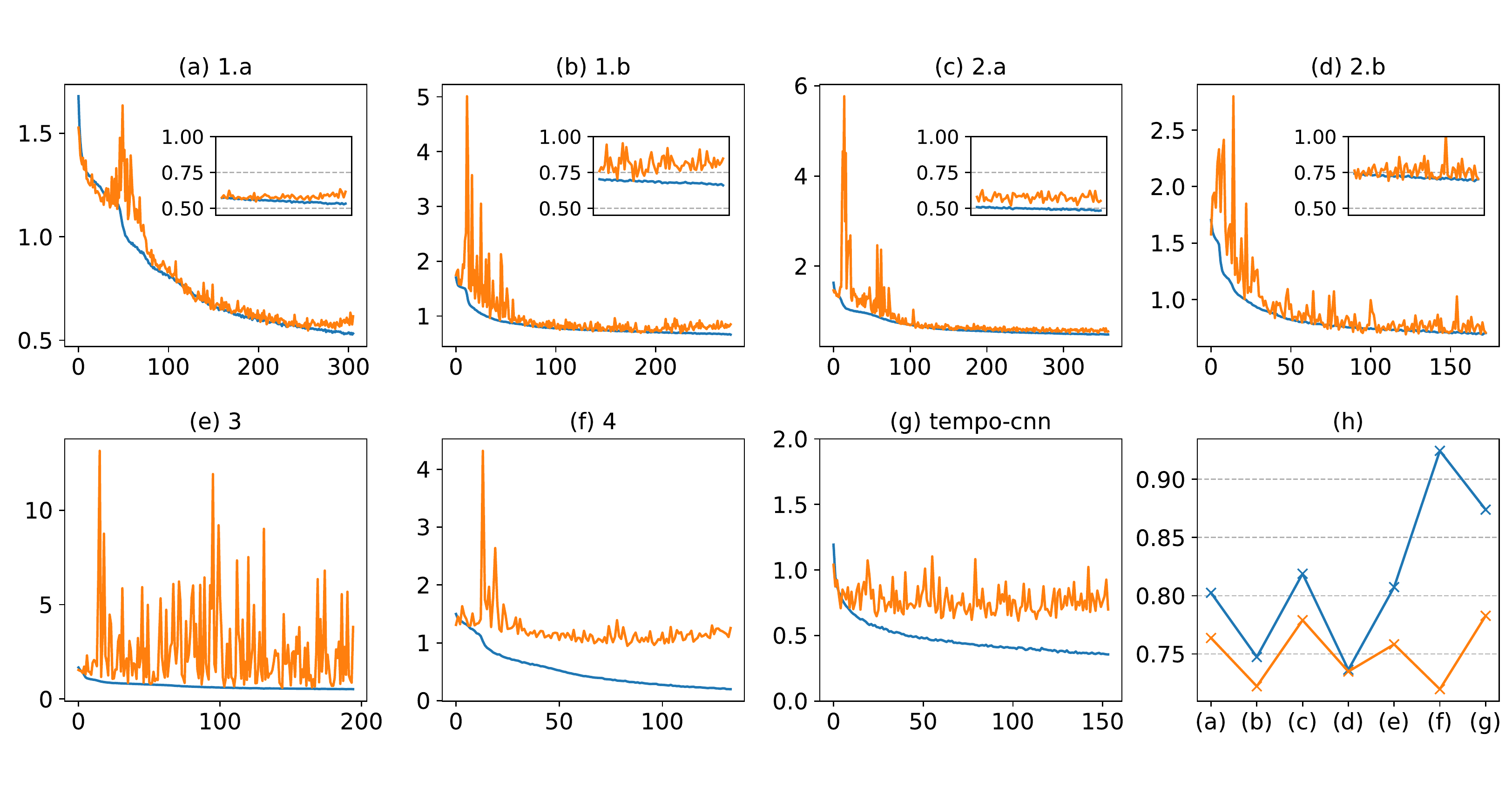}
 \caption{(a) - (g) Training (in blue) and validation (in orange) loss curves for the different architecture variations (inset: the last 70 epochs). (h) Training and validation accuracies at the minimum validation loss for each model. Model labels and their descriptions appear in Table \ref{tab:model_arch_alt}.}
\label{fig:loss_curves}
\end{figure}

We first investigate the training and validation loss curves for all the model variants listed in Table~\ref{tab:model_arch_alt} to justify the design choices.  These appear in Figure~\ref{fig:loss_curves}, with the last 70 epochs of the training shown in the inset plot (only for (a) - (d)). Since the training was halted if validation loss did not reduce for 50 epochs, the last 70 epochs capture the point of lowest loss and the region around it. Taking a closer look at these can help compare the lowest loss values obtained and the stability of each model.\\ 

What we typically would like at the end of a training session is that the training and validation losses be close to each other in value and monotonically decrease (converge) to a stable loss value \cite{brownlee18}. This helps ascertain that the model is able to learn and generalise well without overfitting, and that successive training sessions are likely to give similar results since the validation performance is consistent across epochs. We therefore expect the use of dropout and a model architecture that contains more weights in the convolutional layers (feature learning layers) than in the final dense layers (classification layers), to help. Further, given the small number of output classes in this task, we can perhaps also expect that increasing the model capacity by using more filters of larger kernel sizes in the MF modules (like in tempo-cnn) is unlikely to bring significant improvement. With regards to all these points, the following observations can be made from the loss curves of Figure~\ref{fig:loss_curves}:
\begin{itemize}
    \item Increasing depth by using more MF modules is not of benefit in the case of either model 1 or 2. This is reflected in the higher minimum validation loss in plots (b) and (d), and correspondingly lower accuracy values in plot (h).
    \item A slightly lower loss, higher accuracy, and better convergence in both training and validation is achieved with model 2.a when compared to the rest.
    \item Without the early dropout layers, model 3 overfits quite badly, as the validation loss changes unpredictably across epochs.
    \item Without the average pooling layer at the end, in model 4, the number of weights in the dense layer is much higher than the rest of the network put together, and the model again overfits. The training accuracy achieved is the highest, but the validation accuracy is the lowest among all.
    \item Similarly, with the tempo-cnn architecture, the training loss keeps decreasing but the validation loss decreases consistently mostly only in the first 25-30 epochs, after which it is quite unstable and the model overfits.
    \item Although the tempo-cnn architecture achieves a slightly higher validation accuracy than 2.a, its convergence is poor and therefore results obtained in successive runs are likely to be quite different. 
\end{itemize}

We next look at the average 3-fold cross-validation accuracy values achieved with the proposed model architecture. This accuracy, similar to \textit{accuracy0} in \cite{schreiber-ismir18} measures the proportion of 8s examples for which the s.t.m. was correctly identified. The model used to obtain predictions on the validation set in each fold is obtained from the epoch during the training with the least validation loss. Table~\ref{tab:acc0} shows the results for all three cases - estimation of net s.t.m. from the original mixture, and that of the individual instruments from the separated audios.
\begin{table}
\centering
\begin{tabular}{@{}llll@{}}
\toprule
Case          & Net s.t.m & Vocal s.t.m & Pakhawaj s.t.m \\ \midrule
Accuracy  & 75.23     & 69.07       & 76.84          \\ \bottomrule
\end{tabular}
\caption{Average 3-fold cross-validation accuracies (\%) for surface tempo multiple estimation.}
\label{tab:acc0}
\end{table}
The results are poorer for separated vocals and better for pakhawaj, which reflects also in the net score, because the net s.t.m is dominated by that of the pakhawaj. The class-wise performance is shown using a confusion matrix for each case in Table~\ref{tab:confmats}. In the case of vocals, the classes 1 and 8 are estimated more accurately than the other two . For class 8, this could be due to the distinct nature of vocalisation and the limited diversity of examples due to fewer available sections. For class 1, most examples come from sections where the bandish is sung at a steady rate without improvisation thus making tempo estimation easier. For class 2, sections often come from the earlier stages of improvisation in a concert where the singing is not fully rhythmic and is characterized by pauses, melismatic singing and changes to other s.t.m. levels, making the estimation harder. The confusions between classes 2 and 4 could also be due to some bleed of pakahwaj into the vocals during source separation.\\

\begin{table}
\centering
\begin{tabular}{@{}lcllllll@{}}
\cmidrule(l){4-8}
    & \multicolumn{1}{l}{}                          &                         & \multicolumn{5}{c}{Predicted}         \\ \cmidrule(l){4-8} 
    & \multicolumn{1}{l}{}                          &                         & 1 & 2                    & 4 & 8 & 16 \\ \cmidrule(l){4-8} 
    & \multicolumn{1}{c|}{\multirow{16}{*}{\rotatebox[origin=c]{90}{Ground truth}}} & \multicolumn{1}{l|}{1}  & \textbf{90.15} & 2.17  & 6.38  & 0.0   & 1.3    \\
    & \multicolumn{1}{c|}{}                         & \multicolumn{1}{l|}{2}  & 5.83  & \textbf{82.03} & \textbf{10.85} & 0.79  & 0.50  \\
(a) & \multicolumn{1}{c|}{}                         & \multicolumn{1}{l|}{4}  & 4.51  & \textbf{13.36} & \textbf{66.92} & \textbf{11.93} & 3.28  \\
    & \multicolumn{1}{c|}{}                         & \multicolumn{1}{l|}{8} & 2.43  & 1.81  & \textbf{14.36} & \textbf{65.66} & \textbf{15.74} \\
    & \multicolumn{1}{c|}{}                         & \multicolumn{1}{l|}{16} & 1.84  & 1.01  & 6.48  & \textbf{15.14} & \textbf{75.53} \\
    & \multicolumn{1}{c}{}                         &                         &   & \multicolumn{1}{c}{} &   &   &    \\
    & \multicolumn{1}{c|}{}                         & \multicolumn{1}{l|}{1}  & \textbf{77.30} & \textbf{15.47} & 5.26 & 1.97        &    \\
(b) & \multicolumn{1}{c|}{}                         & \multicolumn{1}{l|}{2}  & \textbf{21.00} & \textbf{50.78} & \textbf{26.20} & 2.02   &    \\
    & \multicolumn{1}{c|}{}                         & \multicolumn{1}{l|}{4}   & 5.84           & \textbf{20.15} & \textbf{64.60} & 9.41   &    \\
    & \multicolumn{1}{c|}{}                         & \multicolumn{1}{l|}{8}  & 1.84           & 0.05           & \textbf{13.25} & \textbf{84.86}  &    \\
    & \multicolumn{1}{c}{}                         &                         &   & \multicolumn{1}{c}{} &   &   &    \\
    & \multicolumn{1}{c|}{}                         & \multicolumn{1}{l|}{1}  & \textbf{92.99} & 0.94 & 5.01 & 0.81 & 0.25 \\
    & \multicolumn{1}{c|}{}                         & \multicolumn{1}{l|}{2}  & 0.20 & \textbf{83.34} & \textbf{14.42} & 1.69           & 0.35           \\
(c) & \multicolumn{1}{c|}{}                         & \multicolumn{1}{l|}{4}  & 5.78 & \textbf{15.11} & \textbf{65.41} & \textbf{11.13} & 2.57           \\
    & \multicolumn{1}{c|}{}                         & \multicolumn{1}{l|}{8} & 2.95 & 1.15           & \textbf{11.48} & \textbf{69.95} & \textbf{14.47} \\
    & \multicolumn{1}{c|}{}                         & \multicolumn{1}{l|}{16} & 1.45 & 1.07           & 6.11           & \textbf{24.77} & \textbf{66.60} \\
    & \multicolumn{1}{l}{}                          &                         &   & \multicolumn{1}{c}{} &   &   &   
\end{tabular}
\caption{Confusion matrix of (a) net, (b) vocal, and (c) pakhawaj s.t.m. predictions (values in \%).}
\label{tab:confmats}
\end{table}

In the case of net and pakhawaj s.t.m., classes 1 and 2 are estimated quite accurately, while the other classes are confused with their immediate neighbours. The class 16 being confused with 8 is most likely because of the presence of accents on every other stroke. We also notice a drop in the performance of this class in the case of separated pakhawaj when compared to the mixture audios, possibly due to a further loss of faint onsets after separation.

\subsection{Boundary Detection and Section Labelling}
We evaluate boundary retrieval performance using the precision, recall and F-score values. A predicted boundary is declared a hit if it falls within a certain duration of an unmatched ground truth boundary, and a false alarm otherwise. Results are reported at two values of temporal tolerance: $\pm$1.5s and $\pm$3s. The latter value is as used in \cite{ulrich14} and the former is included with the reason that given the large number of sections that are 6-9s long, even if both the detected boundaries are off, the detected section still captures at least half of the ground truth section. These results are shown in Table~\ref{tab:bound_det} using both the methods \texttt{seg1} \& \texttt{seg2}.\\

To evaluate section labelling, we report labelling accuracy as the fraction of the duration of each concert that is correctly labelled (excluding regions where the ground truth is not one of \{1,2,4,8,16\}), averaged across the dataset, as defined in \cite{paulus_labeleval}. Each of the three s.t.m. labels are first evaluated individually and also when taken together (i.e., a region is said to be correctly labelled only if all three labels are correct). The results appear in Table~\ref{tab:seclabel_acc}. We expect these scores to be different from the cross-validation accuracies reported in Table~\ref{tab:acc0} as the test set is now no longer balanced. An additional challenge is that some of the confused classes also happen to be the more common ones.\\

\begin{table}
\begin{subtable}[h]{\linewidth}
\centering
\begin{tabular}{@{}lllllll@{}}
\toprule
 & \multicolumn{3}{c}{\begin{tabular}[c]{@{}c@{}}$\pm$1.5s tolerance\end{tabular}} & \multicolumn{3}{c}{\begin{tabular}[c]{@{}c@{}}$\pm$3s tolerance\end{tabular}} \\ \midrule
                                                                     & Prec. & Rec. & F-sc.        & Prec. & Rec. & F-sc.        \\ \midrule
\begin{tabular}[c]{@{}l@{}}\texttt{seg1}\end{tabular} & 0.135     & 0.758  & 0.229           & 0.161     & 0.899  & 0.273          \\
\begin{tabular}[c]{@{}l@{}}\texttt{seg1+}\end{tabular}           & 0.270      & 0.377  & 0.315          & 0.390      & 0.543  & 0.454          \\
\begin{tabular}[c]{@{}l@{}}\texttt{seg2}\end{tabular}  & 0.140      & 0.734  & 0.235          & 0.165     & 0.877  & 0.278          \\
\begin{tabular}[c]{@{}l@{}}\texttt{seg2+}\end{tabular}           & 0.290      & 0.377  & \textbf{0.328} & 0.398     & 0.527  & 0.454 \\ \bottomrule
\end{tabular}
\caption{}
\label{tab:bound_det}
\end{subtable}

\begin{subtable}[h]{\linewidth}
\centering
\begin{tabular}{@{}llllll@{}}
\toprule
 &
  \begin{tabular}[c]{@{}l@{}}Vocals\end{tabular} &
  \begin{tabular}[c]{@{}l@{}}Pakhawaj\end{tabular} &
  \begin{tabular}[c]{@{}l@{}}Net from \\ model\end{tabular} &
  \begin{tabular}[c]{@{}l@{}}Net as \\ max.\end{tabular} &
  \begin{tabular}[c]{@{}l@{}}All 3 \\ labels\end{tabular} \\ \midrule
\begin{tabular}[c]{@{}l@{}}\texttt{seg1} \end{tabular} & 65.42          & 66.72          & 64.54          & 65.51 & 44.19          \\
\begin{tabular}[c]{@{}l@{}}\texttt{seg1+}\end{tabular}           & 67.18          & 68.68          & 66.86          & 67.48 & 45.94          \\
\begin{tabular}[c]{@{}l@{}}\texttt{seg2}\end{tabular}  & 65.42          & 68.90          & 68.58          & -     & 45.99          \\
\begin{tabular}[c]{@{}l@{}}\texttt{seg2+}\end{tabular}           & \textbf{67.72} & \textbf{71.05} & \textbf{70.39} & -     & \textbf{48.61} \\ \bottomrule
\end{tabular}
\caption{}
\label{tab:seclabel_acc}
\end{subtable}
\caption{(a) Boundary detection performance and (b) s.t.m. labelling accuracies (in \%).}
\end{table}

\begin{figure}[!h]
\centering
\begin{subfigure}{0.45\linewidth}
\centering
\includegraphics[trim=10 0 450 30, clip, scale=0.4]{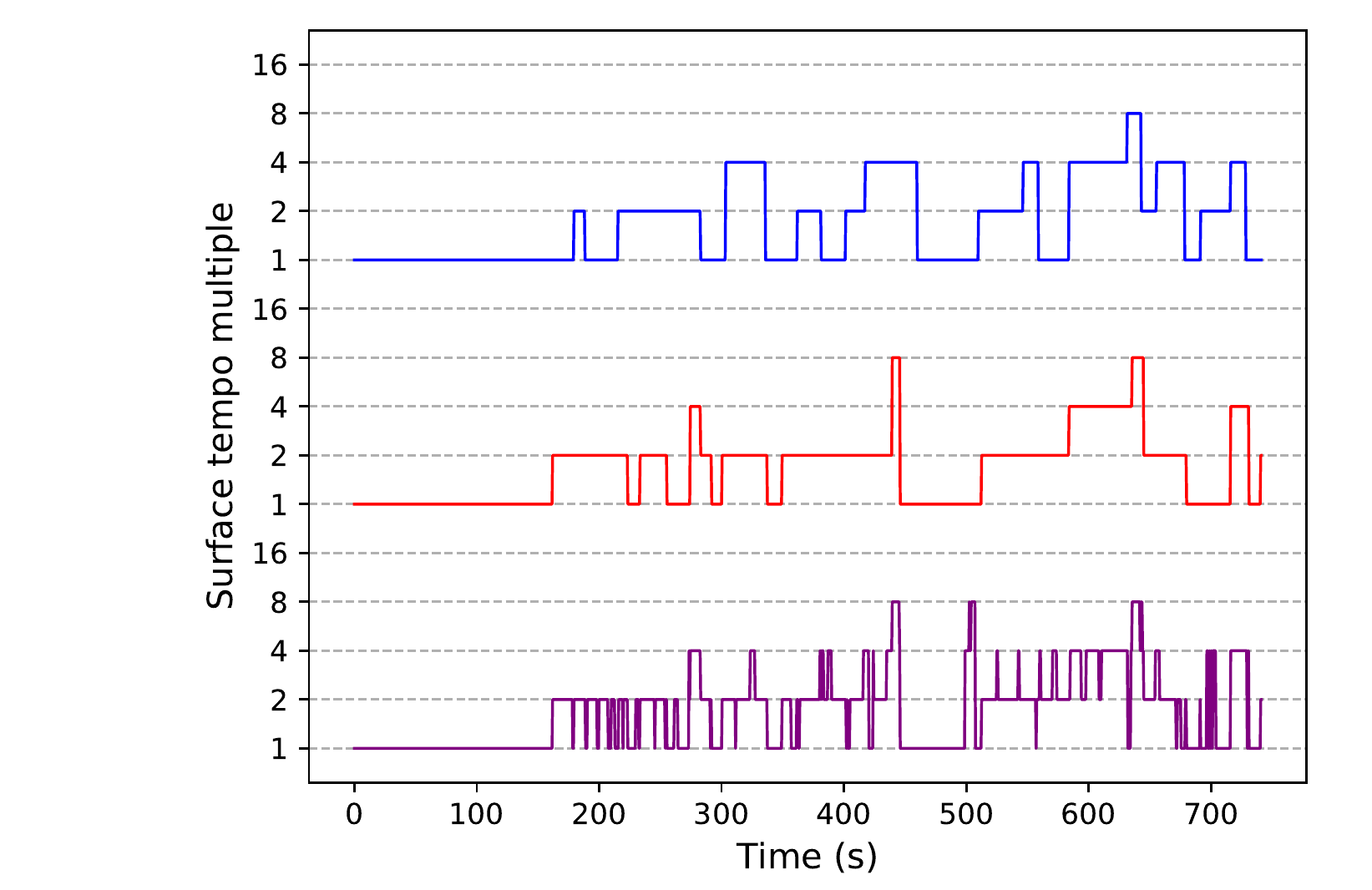}
\caption{}
\label{fig:voc_stm_test}
\end{subfigure}
~~~~~~
\begin{subfigure}{0.45\linewidth}
\centering
\includegraphics[trim=470 0 10 30, clip, scale=0.4]{figures/GBAhirBhrvChoutal_tempo_pred_smu.pdf}
\caption{}
\label{fig:pakh_stm_test}
\end{subfigure}
\caption{The estimated s.t.m. - without (bottom) \& with smoothing (center), and the ground truth (top) s.t.m. labels for (a) vocals and (b) pakhawaj, across the concert \textit{GB\_AhirBhrv\_Choutal}.}
\label{fig:test_stm}
\end{figure}

The individual labelling accuracies are quite similar for the pakhawaj and net tempo labels, slightly lower for the vocals, but much lower for getting all the labels right in every frame. With \texttt{seg1}, we see that the vocal and pakhawaj estimates are reliable enough that taking their maximum as the net s.t.m. instead of using the model estimate improves the net s.t.m. labelling accuracy. Hence, for the evaluation in the last column, the net is taken as the maximum of the other two. Although this seemingly renders the model trained to predict the net s.t.m. not very useful, we see in \texttt{seg2+} that using it to obtain all the estimates together, followed by smoothing, improves all the accuracies, proving its utility.\\

Although better tempo estimation should result in better boundary detection since the boundaries are based entirely on tempo changes, the boundary detection results using \texttt{seg2} are only slightly better than \texttt{seg1}. In both the cases, the smoothing step, as expected, proves useful but significantly reduces the recall. Similar results are observed at both values of temporal tolerance with the higher tolerance, not surprisingly, improving the overall performance. Looking at the vocal and pakhawaj s.t.m. estimates obtained using \texttt{seg2+} in Figure~\ref{fig:test_stm}, we see that for both the instruments, at a coarse level, the various surface tempo regions are captured well. And while for the pakhawaj, finer section changes are also estimated accurately, such changes are not tracked well in the case of vocals, thus reducing the overall boundary detection scores.

\section{Conclusions and Future Work}
We have presented a system that provides a complete rhythmic description of a Dhrupad bandish performance, enabling its segmentation into musicologically relevant sections based on the rhythmic interaction between the instruments. The metric tempo is estimated by adapting existing methods whereas the surface tempo is estimated using deep learning on a spectro-temporal representation of the audio in a novel manner by predicting its relationship with the metric tempo to directly obtain the musically significant \textit{lay} ratio. 
We apply the model to source separated audio streams corresponding to vocal and pakhawaj and because of the challenges presented by imperfect source separation, we benefit from using a model trained also on the mixture audios. We find that the surface tempo multiples at the lower and higher extremes are estimated better than the intermediate values. This, despite the intermediate values being the more represented classes in the dataset, points to the diversity in the acoustic realisations of the different surface densities. That is, the lowest and highest densities are probably also being recognised by the distinct voice/instrument styles as captured by the learned representation. The use of a post-processing step imposing a constraint on the section duration helps by improving the tempo labelling accuracy and the boundary detection performance. This work therefore provides an example of adapting available MIR methods to music genre specific problems.\\

The occurrence of more confusions in the intermediate s.t.m. classes could also be because of cases where an example could very well belong to two s.t.m. classes depending on its m.t. value (e.g., a s.t. value of 140 BPM is 2 times a m.t. of 70 BPM but 4 times a m.t. of 35 BPM). While the fairly narrow metric tempo range of the dataset in the present work might have reduced the number of such cases, there would still have been a few, given that the range spans a little over an octave. Hence, when tested on audios with m.t. outside this range, the models trained on this dataset with the proposed architecture might produce more errors in predicting the intermediate s.t.m. classes. However, it is also likely that the lowest and highest s.t.m. values might still be predicted accurately because of the associated distinct timbres which might have got learned by the model. \\

Future extensions to the present work could thus involve devising ways to jointly estimate the metric tempo and surface tempo multiple. This could be achieved with a multi-task learning approach with separate classification branches built on top of the same feature learning layers. More data would however need to be used to address the class imbalance for the metric tempo estimation task. Source separation could also be improved by introducing new loss functions that preserve onsets better and hence allow better tempo estimation on separated audios. Finally, the surface tempo estimated using these methods can be used along with other cues in the higher-level task of labelling a section improvised or un-improvised. The additional cues could be based on trying to find occurrences of melodic motifs that match the refrain, or of pakhawaj stroke sequences that match the theka, to identify when each instrument is improvising and not.

\pagebreak
\bibliographystyle{plain}
\bibliography{refers}

\begin{thebibliography}{10}

\bibitem{brownlee18}
J.~Brownlee.
\newblock {\em Better Deep Learning: Train Faster, Reduce Overfitting, and Make
  Better Predictions}.
\newblock Machine Learning Mastery, 2018.

\bibitem{mtl-beat-tempo-19}
S.~Böck, M.~E.~P. Davies, and P.~Knees.
\newblock Multi-task learning of tempo and beat: Learning one to improve the
  other.
\newblock In {\em Proc. of the International Society for Music Information
  Retrieval Conference}, 2019.

\bibitem{chordia09}
P.~Chordia and A.~Rae.
\newblock Using source separation to improve tempo detection.
\newblock In {\em Proc. of the International Society for Music Information
  Retrieval Conference}, 2009.

\bibitem{clayton}
M.~Clayton.
\newblock {\em Time in Indian Music: Rhythm, Metre, and Form in North Indian
  Rāg Performance, Volume 1}.
\newblock Oxford University Press, 2000.

\bibitem{elowson13}
A.~Elowsson, A.~Friberg, G.~Madison, and J.~Paulin.
\newblock Modelling the speed of music using features from harmonic/percussive
  separated audio.
\newblock In {\em Proc. of the International Society for Music Information
  Retrieval Conference}, 2013.

\bibitem{aggelos12}
A.~{Gkiokas}, V.~{Katsouros}, G.~{Carayannis}, and T.~{Stajylakis}.
\newblock Music tempo estimation and beat tracking by applying source
  separation and metrical relations.
\newblock In {\em Proc. of IEEE International Conference on Acoustics, Speech
  and Signal Processing}, 2012.

\bibitem{grosche09}
P.~Grosche and M.~Müller.
\newblock A mid-level representation for capturing dominant tempo and pulse
  information in music recordings.
\newblock In {\em Proc. of the International Society for Music Information
  Retrieval Conference}, 2009.

\bibitem{spleeter}
R.~Hennequin, A.~Khlif, F.~Voituret, and M.~Moussallam.
\newblock Spleeter: A fast and state-of-the art music source separation tool
  with pre-trained models.
\newblock Late-Breaking/Demo International Society for Music Information
  Retrieval Conference, 2019.
\newblock Deezer Research.

\bibitem{klapuri05}
A.~P. Klapuri and J.~T.~Astola A.~J.~Eronen.
\newblock Analysis of the meter of acoustic musical signals.
\newblock {\em IEEE Transactions on Audio, Speech, and Language Processing},
  14(1), 2006.

\bibitem{vowel-onset-det}
P.~Kumar, M.~Joshi, S.~Hariharan, S.~Dutta-Roy, and P.~Rao.
\newblock Sung note segmentation for a query-by-humming system.
\newblock In {\em Proc. of Music-AI (International Workshop on Artificial
  Intelligence and Music) in the International Joint Conferences on Artificial
  Intelligence}, 2007.

\bibitem{olivier19}
L.~Olivier and G.~Didier.
\newblock Tempo and metrical analysis by tracking multiple metrical levels
  using autocorrelation.
\newblock {\em Applied Sciences}, 9(23), 2019.

\bibitem{paulus_labeleval}
J.~Paulus and A.~Klapuri.
\newblock Music structure analysis using a probabilistic fitness measure and a
  greedy search algorithm.
\newblock {\em IEEE Transactions on Audio, Speech, and Language Processing},
  17(6), 2009.

\bibitem{peeters07}
G.~Peeters.
\newblock Template-based estimation of time-varying tempo.
\newblock {\em EURASIP Journal on Advances in Signal Processing}, 1(14), 2007.

\bibitem{pr-tismir20}
P.~Rao, T.~P. Vinutha, and M.~A. Rohit.
\newblock Structural segmentation of alap in \uppercase{D}hrupad vocal
  concerts.
\newblock In {\em Transactions of the International Society for Music
  Information Retrieval, Under review}, 2020.

\bibitem{ross-ismir12}
J.~C. Ross, T.~P. Vinutha, and P.~Rao.
\newblock Detecting melodic motifs from audio for \uppercase{H}industani
  classical music.
\newblock In {\em Proc. of the International Society for Music Information
  Retrieval Conference}, 2012.

\bibitem{rubberband}
{Rubber Band Library}.
\newblock Rubber band library v1.8.2.
\newblock \url{https://breakfastquay.com/rubberband/}, 2018.

\bibitem{schreiber-ismir18}
H.~Schreiber and M.~Müller.
\newblock A single-step approach to musical tempo estimation using a
  convolutional neural network.
\newblock In {\em Proc. of the International Society for Music Information
  Retrieval Conference}, 2018.

\bibitem{schreiber-smc19}
H.~Schreiber and M.~Müller.
\newblock Musical tempo and key estimation using convolutional neural networks
  with directional filters.
\newblock In {\em Proc. of the Sound and Music Computing Conference}, 2019.

\bibitem{dunya}
A.~Srinivasamurthy, G.~K. Koduri, S.~Gulati, V.~Ishwar, and X.~Serra.
\newblock Corpora for music information research in \uppercase{I}ndian art
  music.
\newblock In {\em A. Georgaki and G. Kouroupetroglou eds., Proc. of the
  International Computer Music Conference}. Michigan Publishing, 2014.

\bibitem{ajay14}
A.~{Srinivasamurthy} and X.~{Serra}.
\newblock A supervised approach to hierarchical metrical cycle tracking from
  audio music recordings.
\newblock In {\em Proc. of IEEE International Conference on Acoustics, Speech
  and Signal Processing}, 2014.

\bibitem{ulrich14}
K.~Ullrich, J.~Schlüter, and T.~Grill.
\newblock Boundary detection in music structure analysis using convolutional
  neural networks.
\newblock In {\em Proc. of the International Society for Music Information
  Retrieval Conference}, 2014.

\bibitem{verma-icassp15}
P.~Verma, T.P. Vinutha, P.~Pandit, and P.~Rao.
\newblock Structural segmentation of hindustani concert audio with posterior
  features.
\newblock In {\em Proc. of IEEE International Conference on Acoustics, Speech
  and Signal Processing}, 2015.

\bibitem{vinutha-ismir16}
T.~P. Vinutha, S.~Sankagiri, K.~K. Ganguli, and P.~Rao.
\newblock Structural segmentation and visualization of sitar and sarod concert
  audio.
\newblock In {\em Proc. of the International Society for Music Information
  Retrieval Conference}, 2016.

\bibitem{vinutha-ncc16}
T.~P. Vinutha, S.~Sankagiri, and P.~Rao.
\newblock Reliable tempo detection for structural segmentation in sarod
  concerts.
\newblock In {\em Proc. of the National Conference on Communications}, 2016.

\bibitem{bonnie_wade}
B.~C. Wade.
\newblock {\em Music in India: The classical traditions}.
\newblock Prentice-Hall, 1979.

\bibitem{fhfwu11}
F.~H.~F. Wu, T.~C. Lee, J.~S.~R. Jang, K.~K. Chang, C.~H. Lu, and W.~N. Wang.
\newblock A two-fold dynamic programming approach to beat tracking for audio
  music with time-varying tempo.
\newblock In {\em Proc. of the International Society for Music Information
  Retrieval Conference}, 2011.

\end{thebibliography}

\pagebreak
\appendix
\section{Dataset}

Artists:
\begin{itemize}
    \item GB: Gundecha Brothers
    \item UB: Uday Bhawalkar
    \item RFD: R. Fahimuddin Dagar
\end{itemize}

Concert naming convention: artist\_raga\_tala.\\

\begin{table}[hbt!]
\begin{tabular}{@{}lllll@{}}
\toprule
& Concert name           & Source                                                         & m.t. range                                                       & Net s.t. range     \\ \midrule
1 & GB\_AhirBhrv\_Choutal  & MusicBrainz: \href{https://musicbrainz.org/recording/178b4cf6-88e6-414d-bfbd-3d90bb368a9a}{\underline{Link}} & 50 - 65 & 50 - 947  \\
2 & GB\_Bhim\_Choutal      & MusicBrainz: \href{https://musicbrainz.org/recording/b5187674-c65e-4ce5-8e0a-5be891fc9bcb}{\underline{Link}} & 49 - 61 & 98 - 961  \\
3 & GB\_BKT\_Choutal       & MusicBrainz: \href{https://musicbrainz.org/recording/d4aa21db-fddc-4fdf-8d09-c0a4fcf9066a}{\underline{Link}} & 47 - 55 & 95 - 443  \\
4 & GB\_Jai\_Choutal       & YouTube: \href{https://youtu.be/ttBbyqt8C28?t=1195}{\underline{Link}}                & 44 - 50 & 45 - 795  \\
5 & GB\_MMal\_Choutal      & MusicBrainz: \href{https://musicbrainz.org/recording/d3bbce17-0af9-47fd-9420-9d0807201774}{\underline{Link}} & 47 - 53   & 51 - 421  \\
6 & GB\_Marwa\_Choutal     & MusicBrainz: \href{https://musicbrainz.org/recording/68292723-7f44-4da4-9aa9-2308eb3ec9ee}{\underline{Link}} & 46 - 62 & 110 - 941  \\
7 & GB\_Bhrv\_Choutal      & MusicBrainz: \href{https://musicbrainz.org/recording/59c88c32-0bde-433b-b194-0f65281e5714}{\underline{Link}} & 45 - 56 & 181 - 884  \\
8 & GB\_Bhg\_Dhamar        & MusicBrainz: \href{https://musicbrainz.org/recording/3b7489cf-2106-4bd2-99b8-9e886d945ddf}{\underline{Link}} & 48 - 60 & 48 - 895   \\
9 & GB\_Kedar\_Dhamar      & YouTube: \href{https://www.youtube.com/watch?v=NlOjCn0MVWI}{\underline{Link}}        & 45 - 52 & 179 - 412   \\
10 & UB\_AhirBhrv\_Tivratal & MusicBrainz: \href{https://musicbrainz.org/recording/24ebe543-4197-4f7e-9be7-d6db7d1c4f2c}{\underline{Link}} & 51 - 65 & 51 - 488 \\
11 & UB\_Bhrv\_Sooltal      & YouTube: \href{https://youtu.be/jG21qUpkfUI?t=3898}{\underline{Link}}                & 43 - 59 & 43 - 889  \\
12 & UB\_Bhg\_Dhamar        & YouTube: \href{https://youtu.be/0PljerNT1iU?t=3115}{\underline{Link}}                & 35 - 50 & 35 - 397   \\
13 & UB\_Shree\_Dhamar      & YouTube: \href{https://youtu.be/NCmvttDzkzk?t=1215}{\underline{Link}}                & 46 - 60 & 185 - 949   \\
14 & UB\_Malkauns\_Choutal  & YouTube: \href{https://www.youtube.com/watch?v=ZLvtqLwO-2Y}{\underline{Link}}        & 47 - 60 & 47 - 823  \\ \midrule \\
\multicolumn{4}{c}{Other concerts, parts of which were used for data augmentation}                                                                                       \\ \midrule
1 & GB\_KRA\_Dhamar &
  MusicBrainz: \href{https://musicbrainz.org/recording/e454b311-5f6d-4f4c-a2ee-a5e8d1d24ea4}{\underline{Link}} & 45 - 52 & 45 - 816 \\
2 & GB\_Rageshri\_Choutal  & YouTube: \href{https://youtu.be/itI7s4TEPAI?t=750}{\underline{Link}}                 & 43 - 64 & 43 - 994  \\
3 & GB\_Yaman\_Choutal     & MusicBrainz: \href{https://musicbrainz.org/recording/5e36d2fc-cddf-4fc2-b336-be537764dbd5}{\underline{Link}} & 44 - 54 & 44 - 54  \\
4 & GB\_DK\_Dhamar         & YouTube: \href{https://youtu.be/jTRKzUT0oHI?t=2856}{\underline{Link}}                & 48 - 59 & 48 - 114   \\
5 & UB\_Lalit\_Dhamar      & YouTube: \href{https://youtu.be/DF2m3ExOCyo?t=2929}{\underline{Link}}                & 35 - 39 & 35 - 630   \\
6 & UB\_Maru\_Choutal      & YouTube: \href{https://youtu.be/AiLEB6e4zvY?t=3497}{\underline{Link}}                & 38 - 49 & 38 - 781  \\
7 & \begin{tabular}[c]{@{}l@{}}RFD\_Kedar\_Choutal, \\ RFD\_Kedar\_Dhamar\end{tabular} &
  MusicBrainz: \href{https://musicbrainz.org/recording/aa45d34e-0ba8-43f5-8ec9-48c50305067a}{\underline{Link}} &
  \begin{tabular}[c]{@{}l@{}}49 - 57, \\ 66 - 74\end{tabular} &
  \begin{tabular}[c]{@{}l@{}}49 - 57, \\ 66 - 74\end{tabular} \\ \bottomrule
\end{tabular}
\caption{List of the audios in the dataset, their sources, and the metric and net surface tempo ranges (in BPM) in each (for the last 7 concerts these ranges are based on the annotated portions only). All the concerts are in the vilambit or madhya lay. The dataset has a m.t. range of 35-75 BPM and a net s.t. range of 35-994 BPM.}
\label{tab:dataset_sources}
\end{table}

\end{document}